\documentclass[final,5p,authoryear,10pt]{elsarticle}

\usepackage{amsmath}
\usepackage{graphicx}
\usepackage{booktabs}

\usepackage{amssymb}
\usepackage{amsthm}
\usepackage{xfrac}
\usepackage{float}
\usepackage{hyperref}
\usepackage{fancyvrb}
\usepackage{color}
\usepackage{footnote} 
\usepackage[usenames,dvipsnames,svgnames,table]{xcolor}
\usepackage{subfig}

\newcommand{\ba}{\begin{eqnarray}}
\newcommand{\ea}{\end{eqnarray}}
\newcommand{\nn}{\nonumber}
\newcommand{\mb}[1]{\mathbf{#1}}
\newcommand{\bs}[1]{\boldsymbol{#1}}

\newcommand{\red}[1]{{\color{black}#1}}  
\journal{Annals of Nuclear Energy}

\begin{document}

\begin{frontmatter}



\title{MOCABA: a general Monte Carlo-Bayes procedure for improved predictions of integral functions of nuclear data}

\author[OFF]{A.~Hoefer}
\ead{axel.hoefer@areva.com}

\author[OFF]{O.~Buss}
\author[OFF]{M.~Hennebach}
\author[OFF]{M.~Schmid}
\author[ERL]{D. Porsch}

\address[OFF]{AREVA~GmbH -- Dep.~Radiology \& Criticality\\
    Kaiserleistrasse 29, 63067 Offenbach am Main, Germany\\}
\address[ERL]{AREVA~GmbH -- Dep.~Neutronics\\
        Paul-Gossen-Strasse 100, 91052 Erlangen, Germany\\}

\begin{abstract}
MOCABA is a combination of Monte Carlo sampling and Bayesian updating algorithms for the 
prediction of integral functions of nuclear data, such as reactor power distributions or 
neutron multiplication factors. Similarly to the established Generalized Linear Least Squares (GLLS) 
methodology, MOCABA offers the capability to utilize integral experimental data to reduce the prior 
uncertainty of integral observables. The MOCABA approach, however, does not involve any series 
expansions and, therefore, does not suffer from the breakdown of first-order 
perturbation theory for large nuclear data uncertainties. 
This is related to the fact that, in contrast to the GLLS method, 
the updating mechanism within MOCABA is applied directly to the integral observables without having to 
``adjust'' any nuclear data. 
A central part of MOCABA is the nuclear data Monte Carlo program NUDUNA, which performs
random sampling of nuclear data evaluations according to their covariance information and 
converts them into libraries for transport code systems like MCNP or 
SCALE. What is special about MOCABA is that it can be applied to any integral function of nuclear data, 
and any integral measurement can be taken into account to improve the prediction of an integral 
observable of interest. In this paper we present two example applications of the MOCABA framework: the prediction of the neutron 
multiplication factor of a water-moderated PWR fuel assembly based on 21 criticality safety benchmark experiments 
and the prediction of the power distribution within a toy model reactor containing 100 fuel assemblies. 
\end{abstract}


\end{frontmatter}


\newpage

%
\section{Introduction}
\label{sect::intro}
%
%
%
Over the last years, methods of Monte Carlo propagation of nuclear data uncertainties have been playing an 
increasingly important role in the uncertainty analysis of integral observables, such as neutron multiplication
factors, isotopic concentrations in irradiated nuclear fuel or reactor power distributions 
\citep{koning_tmc, acab_2008, xsusa, nuduna,xsusa_scale, psi}.
This is largely owed to the fact that comprehensive covariance data have been added to the major nuclear data 
evaluations \citep{endfb71,jendl4,jeff32,talys}. The application range of Monte 
Carlo based nuclear data uncertainty propagation is very wide and extends from criticality safety analysis over reactor core analysis, 
depletion analysis and activation analysis to more exotic applications like the design of accelerator-driven systems and 
fusion systems \citep{nuduna,xsusa_scale,xsusa_core_1,xsusa_transient,acab_buc,acab_ads,rochman_fusion}. 

However, the traditional way of propagating nuclear data uncertainties is {\it not} based on Monte Carlo 
simulation but on adjoint-based first-order perturbation theory, where uncertainties of integral observables due
to nuclear data uncertainties are approximated by linear transformations of nuclear data covariances. 
These linear transformations are defined by the sensitivities of the integral observables to the nuclear data 
\citep{uchasev,gandini,broadhead}. 

A major strength of first-order perturbation theory is that it can be combined with the Generalized Linear Least 
Squares (GLLS) method, which allows us to utilize integral experimental data to 
update prior knowledge about the nuclear data and, consequently, about integral functions 
of the nuclear data \citep{cecchini,humi,hemment,broadhead,saintjean,salvatores}. 
This procedure is also often referred to as ``nuclear data adjustment''. 

In spite of the merits of first-order perturbation theory, there are limitations to its applicability: 
\begin{itemize}
\item
If nuclear data uncertainties are too large, first-order perturbation 
theory breaks down. This may already happen for nuclear data uncertainties in the range of only a few 
percent \citep{gllsm,rochman_tmc}.
\item
Application cases with a very large number of responses, such as time-dependent pin-wise fission rates in a depletion 
analysis, would require an unmanageably large number of adjoint transport calculations \citep{xsusa_scale}.
\item
Many transport codes do not offer the option to perform adjoint calculations. Adjoint-based 
first-order perturbation theory is not applicable in such cases \citep{xsusa_scale}.
\end{itemize}
Fortunately, the above limitations do not apply to the Monte Carlo propagation of nuclear data 
uncertainties. 
An apparent disadvantage of the Monte Carlo approach, however, is that it cannot be combined with
GLLS updating. Hence, what we need is a Bayesian procedure similar to the GLLS method
which allows us to take into account integral benchmark measurements
to update the Monte Carlo predictions of the integral observables of interest. 

Such a Bayesian updating mechanism for Monte Carlo data is the main characteristics of the MOCABA approach. 

The MOCABA procedure is divided into two steps: a Monte Carlo step and an updating step. 
In the Monte Carlo step,
nuclear data and system parameters defining the application case (e.g.~a reactor core) 
are random sampled from their respective uncertainty distributions. 
These random draws are then used as input to transport calculations, which provides a multivariate data 
set of integral observables for the application case and the benchmark experiments. 
In the updating step, information related to the benchmark experiments is added, 
namely the integral measurements as well as the system parameters defining the experimental setups 
and their uncertainties. This provides us with updated estimates and uncertainties 
of the integral observables for the benchmark experiments {\it and} with updated predictions and 
uncertainties of integral observables for the application case. Constraints on linear 
combinations of the integral observables (e.g.~a constraint on the total power of a nuclear reactor) 
may also be included in the updating step.

Depending on the degree of physical similarity between the benchmark experiments and the application 
case\red{\footnote{\red{Here physical similarity is meant in the sense that two systems are similar if their integral observables have 
similar sensitivities to variations in the nuclear data.}}}, the MOCABA approach allows for significantly more precise predictions of 
integral observables than a Monte Carlo approach without Bayesian updating. This makes MOCABA very useful for a lot of applications.

One of the merits of the MOCABA framework is that it can be applied to any 
function of nuclear data, including vectors of local power values defining the power 
distributions within nuclear reactors, 
vectors of isotopic concentrations defining the compositions of irradiated fuel samples, 
or individual neutron multiplication factors. Moreover, any integral measurement 
can be used as a benchmark for the updating procedure. 
We may even include measurements of integral observables that seem very different from those we want 
to predict, like reactor power measurements as benchmarks for the prediction of the isotopic 
composition of a completely unrelated irradiated fuel assembly.
 
In the following, we first present a description of the statistical model 
and its implementation. The MOCABA procedure is then applied to the prediction of the 
neutron multiplication factor of a water-moderated PWR fuel assembly and to the prediction of the power 
distribution within a toy model reactor. We conclude the paper with a short summary and outlook.
%
%
\section{The Bayesian model and its implementation} 
\label{sect::prob_imp}
%
%
%
\subsection{Definition of the Bayesian model}
\label{sect::model}
%
%
%
Let us consider an arbitrary vector function $\mb y$ of a nuclear data vector 
$\bs\alpha$:
\begin{align}
\mb y(\bs\alpha) & \,=\, \left(y_1(\bs\alpha),\dots,y_n(\bs\alpha)\right)^T
,\quad
\bs\alpha \,=\, \left(\alpha_1,\dots,\alpha_r\right)^T .
\label{eq::y}
\end{align}
To express nuclear data uncertainties,
$\bs\alpha$ is treated as a random vector defined by an $r$-variate probability density function (pdf) 
${\rm p}(\bs\alpha)$. 
Consequently, $\mb y(\bs\alpha)$ as a function of $\bs\alpha$ is also a random vector defined by an 
$n$-variate pdf 
${\rm p}(\mb y)$ which reflects the uncertainty of $\mb y$ due to nuclear data uncertainties. 
${\rm p}(\mb y)$ is identified with an $n$-variate normal distribution defined 
by a mean vector $\mb y_0$ and a covariance matrix $\bs\Sigma_0$:\footnote{\red{In case the normality 
assumption does not hold, e.g.~for nonlinear responses of $\mb y$ to variations in the nuclear data $\bs\alpha$,
$\mb y$ may be mapped onto an approximately normally distributed vector $\mb z$ by means of an 
invertible variable transformation $f$. In that way ${\rm p}(\mb y)$ may be chosen from a more general class of distribution models 
such that the response of $\mb y$ to variations in $\bs\alpha$ is correctly reflected. 
The normal Bayesian model described below then applies to the transformed vector $\mb z$, and the distribution of $\mb y$ is obtained 
simply by applying the inverse transformation $f^{-1}$ to $\mb z$.}}
\begin{align}
  {\rm p}(\mb y) & \,=\, {\rm N}\left(\mb y_0,\bs\Sigma_0\right) \,\propto\, \exp\left(-Q_0/2\right), \nn\\
Q_0 & \,=\,\left(\mb y - \mb y_0\right)^T\bs\Sigma_0^{-1}\left(\mb y - \mb y_0\right).
\label{eq::prior_y}
\end{align}
Since ${\rm p}(\mb y)$ reflects our knowledge about $\mb y$ before any measurements of $\mb y$ or 
constraints on $\mb y$ are taken into account, we refer to ${\rm p}(\mb y)$ as the prior pdf of $\mb y$. 

Measurements of $\mb y$ and linear constraints on $\mb y$ may both be expressed in terms of a likelihood function of the following type:
\begin{align}
{\rm p}(\mb v \,|\, \mb y) &
\,\propto\, \exp\left(-Q_V/2\right), \nn\\
\quad Q_V & \,=\,\bs\Delta^T\bs\Sigma_V^{-1}\bs\Delta, \quad \bs\Delta := \mb U\mb y - \mb v .
\label{eq::like_y}
\end{align}
Here $\mb U$ is a rectangular matrix, i.e.~$\mb U \mb y$ represents a linear transformation of $\mb y$, 
and $\mb v$ is a vector defining the measurements and/or linear constraints. Hence, $\mb v$ represents 
the best estimate of $\mb U\mb y$ and $\bs\Sigma_V$ is the corresponding covariance matrix.

According to Bayes' theorem, the updated information about $\mb y$, which includes the
prior knowledge related to the nuclear data as well as the integral measurements and constraints, is 
represented by the posterior pdf, which is defined as the normalized product of the prior pdf 
and the likelihood function \citep{gelman}:
\begin{align}
{\rm p}(\mb y \,|\, \mb v) & \,\propto\, {\rm p}(\mb v\, |\, \mb y) \, {\rm p}(\mb y)\,.
\label{eq::post_y}
\end{align}
Since both ${\rm p}(\mb y)$ and ${\rm p}(\mb v \,|\, \mb y)$ are multivariate normal, 
Eq.~(\ref{eq::post_y}) yields a multivariate normal posterior pdf:
\begin{align}
{\rm p}(\mb y \,|\, \mb v) & \,=\, {\rm N}\left(\mb y^*,\bs\Sigma^* \right) 
\,\propto\, \exp\left(-Q^*/2\right),\nn\\ 
Q^* &\,=\, Q_0+Q_V \,=\,\left(\mb y - \mb y^*\right)^T\bs\Sigma^{*-1}\left(\mb y - \mb y^*\right) .
\label{eq::post_y_normal}
\end{align}
Here $\mb y^*$ is the maximum-a-posteriori estimate and 
$\bs\Sigma^*$ the related posterior covariance matrix of $\mb y$.

To distinguish between observables related to the application case and observables related to the 
benchmark experiments, $\mb y$ may be partitioned into an application case vector $\mb y_A$ 
and a benchmark vector $\mb y_B$:
\begin{align}
  \mb y &\,=\, \left( \mb y_A^T  , \mb y_B^T \right)^T .
\label{eq::partition_y}
\end{align}
Accordingly, we also express the model parameters of the prior distribution, the 
likelihood function and the posterior distribution in partitioned form:
\begin{align}
 \mb y_0 &\,=\, \left( \mb y_{0A}^T  , \mb y_{0B}^T \right)^T
 ,\;\;
 \bs\Sigma_0 \,=\,
 \left(
 \begin{matrix}
 \bs\Sigma_{0A} & \bs\Sigma_{0AB} \\
 \bs\Sigma_{0AB}^T & \bs\Sigma_{0B}
 \end{matrix}
 \right) 
 , \nn\\
 \mb v &\,=\,  \left( \mb v_A^T  , \mb v_B^T \right)^T 
 ,\;\;\;\;
\bs\Sigma_V \,=\,
\left(
\begin{matrix}
\bs\Sigma_{VA} & \mb 0\\
\mb 0^T & \bs\Sigma_{VB}
\end{matrix}
\right),
\nn\\
\mb U &\,=\,
\left(
\begin{matrix}
\mb U_A & \mb 0 \\
\mb 0^T & \mb U_B
\end{matrix}
\right),\;\,
\mb y^* \,=\, \left( \mb y_A^{*T}  , \mb y_B^{*T} \right)^T
, 
\nn\\
\bs\Sigma^* &\,=\,
\left(
\begin{matrix}
\bs\Sigma_A^* & \bs\Sigma_{AB}^* \\
\bs\Sigma_{AB}^{*T} & \bs\Sigma_B^*
\end{matrix}
\right) 
.
\label{eq::partition_prior_like_post}
\end{align}
To obtain the posterior distribution model parameters $\mb y^*$ and $\bs\Sigma^*$, 
the quadratic form $Q^*$ defined in Eqs.~(\ref{eq::prior_y}), (\ref{eq::like_y}) and (\ref{eq::post_y_normal}) has to be
minimized with respect to $\mb y$. For cases with direct measurements of each component of $\mb y_B$ without
constraints, which are represented by
\begin{align}
\bs\Sigma_{VA}^{-1}\,=\,\mb0,\quad \mb U_B\,=\,\mb I,
\label{eq::no_constraints}
\end{align}
we get the following expressions for the posterior model parameters:
\begin{align}
\mb y_A^*  &\,=\,  \mb y_{0A}+\bs\Sigma_{0AB}\left(\bs\Sigma_{0B}+\bs\Sigma_{VB}\right)^{-1}\left(\mb v_B-\mb y_{0B}\right) , \nn\\
\mb y_B^*  &\,=\,  \mb y_{0B}+\bs\Sigma_{0B}\left(\bs\Sigma_{0B}+\bs\Sigma_{VB}\right)^{-1} \left(\mb v_B-\mb y_{0B}\right) ,\nn\\
\bs\Sigma_A^*  &\,=\, \bs\Sigma_{0A}-\bs\Sigma_{0AB}\left(\bs\Sigma_{0B}+\bs\Sigma_{VB}\right)^{-1}\bs\Sigma_{0AB}^T,\nn\\
\bs\Sigma_B^*  &\,=\, \bs\Sigma_{0B}-\bs\Sigma_{0B}\left(\bs\Sigma_{0B}+\bs\Sigma_{VB}\right)^{-1}\bs\Sigma_{0B},\nn\\
\bs\Sigma_{AB}^*  &\,=\, \bs\Sigma_{0AB}-\bs\Sigma_{0AB}\left(\bs\Sigma_{0B}+\bs\Sigma_{VB}\right)^{-1}\bs\Sigma_{0B} \,.
\label{eq::post_model_no_constraints}
\end{align}
As follows from Eq.~(\ref{eq::post_model_no_constraints}), our knowledge about the application case observables $\mb y_A$
can be increased by measurements of the benchmark observables $\mb y_B$, since $\mb y_A$ and $\mb y_B$ are correlated
by common nuclear data uncertainties. These correlations are represented by the submatrix
$\bs\Sigma_{0AB}$ of the prior covariance matrix $\bs\Sigma_0$. 
If the physical characteristics of the application case are very different from those of the benchmark experiments, 
the components of $\bs\Sigma_{0AB}$ tend to be very small. 
Measurements of $\mb y_B$ then only have a very small impact on our knowledge
about $\mb y_A$, i.e.~the posterior estimates $\mb y_A^*$ and 
$\bs\Sigma_A^*$ differ only very little from the prior estimates $\mb y_{0A}$ and $\bs\Sigma_{0A}$. 
Conversely, for benchmark experiments with a very high degree of physical similarity to the application case, 
the correlations corresponding to $\bs\Sigma_{0AB}$ are close to one, in which case 
measurements of the benchmark observables $\mb y_B$ may significantly improve our 
knowledge about $\mb y_A$.

Up to this point, our Bayesian model takes into account nuclear data uncertainties, 
reflected by the prior covariance matrix $\bs\Sigma_0$, as well as uncertainties related to the 
benchmark experiments and constraints, reflected by the likelihood covariance matrix $\bs\Sigma_V$. 
What is still missing are the system parameter uncertainties related to the application case. 
They can, however, be included very easily into the prior pdf ${\rm p}(\mb y)$.
For this purpose, the parameter vector $\mb x_A$ characterizing the system parameters of 
the application case is treated as a random vector defined by a pdf ${\rm p}(\mb x_A)$,
and the integral observable vector $\mb y$ is now considered to be a vector function of 
the nuclear data random vector $\bs\alpha$ {\it and} the system parameter random vector $\mb x_A$:
\begin{align}
  \mb y \,=\, \mb y(\bs\alpha,\mb x_A) \,=\,\left(\mb y_A^T(\bs\alpha,\mb x_A),
 \mb y_B^T(\bs\alpha)\right)^T .
\label{eq::y_app_unc}
\end{align}
Consequently, the prior pdf ${\rm p}(\mb y)$ in Eq.~(\ref{eq::prior_y}) 
is now defined by the pdf ${\rm p}(\bs\alpha)$ reflecting the nuclear data uncertainties 
{\it and} the pdf ${\rm p}(\mb x_A)$ reflecting the system parameter uncertainties of the 
application case. Hence, the prior distribution model parameters $\mb y_0$ and $\bs\Sigma_0$ now 
represent
both the prior information about the nuclear data {\it and} the information about the 
system parameters of the application case. Since the parametric structure of the 
prior pdf is not changed by including system parameter uncertainties of the application case, 
the procedure of calculating the posterior pdf ${\rm p}(\mb y\,|\,\mb v)$ stays the same as 
described above.

Having included system parameter uncertainties of the application case into
our Bayesian model, the list of uncertainties to be considered is complete.
%
%
\subsection{The MOCABA procedure}
\label{sect::mocaba}
%
%
MOCABA is a software implementation of the Bayesian model defined above.

In order to propagate nuclear data uncertainties to integral observable uncertainties, random sampling of 
ENDF-6-formatted nuclear data files is performed with the aid of the nuclear data Monte Carlo program NUDUNA \citep{nuduna} 
using the uncertainty information included in the corresponding covariance files 
of the respective nuclear data evaluation \citep{endfb71,jendl4,jeff32,talys}. \linebreak[4] 
\red{NUDUNA is designed to read input data provided in ENDF-6 format~\cite{endf6_format}. Its random sampling is based on the information provided in the File~31-34 uncertainty sections and in File 8 of an ENDF-6 tape, i.e.~it considers uncertainties of neutron multiplicities, resonance parameters, cross sections, angular distributions, half-lives, and decay branching ratios. Neutron fission spectrum uncertainties, fission yield uncertainties and correlations of different isotopes are not yet included in the current NUDUNA version.} The sampled nuclear data files are subsequently converted automatically into libraries for transport calculations. 

The current version of NUDUNA has the capability to generate random ACE tapes for 
continuous energy transport calculations with MCNP or SERPENT as 
well as 44-group and 238-group AMPX tapes for transport calculations with SCALE \citep{mcnp5,serpent,scale60}. 
To extend the application range to reactor core design and reactor safety analysis, the option 
to generate few-group libraries for reactor core simulation systems will be included in 
future versions of NUDUNA.

To take system parameter uncertainties of the application case into account,
for each random sample $\bs\alpha_i^{MC}$ of the nuclear data evaluation vector 
$\bs\alpha$ generated by NUDUNA, a random sample $\mb x_{A,i}^{MC}$ of the application 
case system parameter vector $\mb x_A$ is drawn. 
The Monte Carlo samples are then used as input parameters for transport calculations 
of the application case and benchmark observables represented by the components of the integral observable vector $\mb y$; 
see Eq.~(\ref{eq::y_app_unc}). Hence, a set of $m$ random draws of $\bs\alpha$ and $\mb x_A$ 
yields a multivariate data set
\begin{align}
  Y_{MC} & \,=\, 
\Bigl\{\mb y_1^{MC},\dots,\mb y_m^{MC}\Bigr\} \nn\\ 
& \,=\,
\Bigl\{\mb y\left(\bs\alpha_1^{MC},\mb x_{A,1}^{MC}\right),\dots,
\mb y\left(\bs\alpha_{m}^{MC},\mb x_{A,m}^{MC}\right)\Bigr\}
\label{eq::Y_MC}
\end{align}
of $m$ random draws of $\mb y$. These data are used for the estimation of the prior pdf ${\rm p}(\mb y)$, where we apply 
the following consistent and unbiased estimators of the prior mean vector and the prior 
covariance matrix \citep{gelman}:
\begin{align}
  \hat{\mb y}_0 &\,=\,\frac{1}{m}\sum_{i=1}^m \mb y_i^{MC} , \nn\\
  \hat{\bs\Sigma}_0 &\,=\,\frac{1}{m-1}\sum_{i=1}^m \left(\mb y_i^{MC}- \hat{\mb y}_0\right)
  \left(\mb y_i^{MC}- \hat{\mb y}_0\right)^T .
\label{eq::MC_estimators}
\end{align}
If a sufficiently large value for the Monte Carlo sample size $m$ is chosen, $\hat{\mb y}_0$ and $\hat{\bs\Sigma}_0$
may be identified with the prior distribution model parameters $\mb y_0$ and $\bs\Sigma_0$; 
see Eq.~(\ref{eq::prior_y}). If the Monte Carlo data $Y_{MC}$ do not fit the assumption 
that $\mb y$ is normally distributed, e.g., if there is a strongly non-linear response 
of $\mb y$ to variations in the nuclear data, we can make use of a suitable invertible variable transformation
$\mb y\to\mb z$ onto an approximately normally distributed random vector $\mb z$ \red{(see footnote 2)}.

Before performing the Bayesian updating according to Eq.~(\ref{eq::post_y}), we have to translate the
benchmark uncertainties into the covariance matrix $\bs\Sigma_V$ of the likelihood function 
${\rm p}(\mb v\,|\,\mb y)$; see Eq.~(\ref{eq::like_y}). 
For direct measurements of the benchmark observables without any constraints (see 
Eq.~(\ref{eq::no_constraints})), the measurement vector $\mb v_B$ is our best estimate of
$\mb y_B$ if we ignore the prior knowledge about the nuclear data, and the covariances of the benchmark measurements 
due to uncertainties in the benchmark system parameters are represented by the covariance matrix $\bs\Sigma_{VB}$; 
see Eq.~(\ref{eq::partition_prior_like_post}).

Collecting the system parameters characterizing all \linebreak benchmark experiments under consideration into a single 
parameter vector $\mb x_B$ allows us to express the benchmark system parameter uncertainties by treating 
$\mb x_B$ as a random vector defined by a pdf ${\rm p}(\mb x_B)$. The benchmark measurements are then expressed as a 
vector function $\mb y_B(\mb x_B)$ of this random vector.
Since we are generally dealing with sets of measurements that are related by common boundary conditions, like 
sets of local reactor power measurements, the benchmark measurements are generally related by common sets 
of uncertain system parameters. Consequently, the components of the random vector $\mb y_B(\mb x_B)$ are generally 
stochastically dependent, and the covariance matrix $\bs\Sigma_{VB}$ is generally not diagonal. 

One way to evaluate the covariance matrix $\bs\Sigma_{VB}$ is via a Monte Carlo method, where we draw a sufficient 
number of random samples of the system parameter vector $\mb x_B$, insert them into transport calculations for $\mb y_B$ and calculate a 
statistical estimate of $\bs\Sigma_{VB}$ analogous to Eq.~(\ref{eq::MC_estimators}); see \cite{buss_physor_2010}.

Another way to calculate $\bs\Sigma_{VB}$ is via a first order series expansion of $\mb y_B$
about the nominal system parameter vector $\mb x_{B,0}$:
\begin{align}
  \mb y_B(\mb x_B) &\,\simeq\, \mb y_B(\mb x_{B,0})+\mb S_B \left(\mb x_B-\mb x_{B,0}\right), 
  \nn\\
\left(\mb S_B\right)_{ij} &\,=\, \left.\partial y_{Bi}/\partial x_{Bj}\right|_{\mb x_B=\mb x_{B,0}},
\label{eq::y_B_expansion}
\end{align}
which is an acceptable approximation if the system parameter uncertainties represented by ${\rm p}(\mb x_B)$ are 
sufficiently small. The covariance matrix $\bs\Sigma_{VB}$ is then obtained by a linear transformation of the 
covariance matrix $\bs\Sigma_{\mb XB}$ of the system parameters $\mb x_B$ defined by the sensitivity matrix $\mb S_B$:
\begin{align}
  \bs\Sigma_{VB} \,\simeq\, \mb S_B \bs\Sigma_{\mb XB} \mb S_B^T.
\label{eq::model_like_linear}
\end{align}
Having calculated the prior model parameters, $\mb y_0$ and $\bs\Sigma_0$, and the covariance matrix of the 
likelihood function $\bs\Sigma_{VB}$, the posterior model parameters, $\mb y^*$ and $\bs\Sigma^*$, are obtained 
by minimizing the quadratic form $Q^*$ in Eq.~(\ref{eq::post_y_normal}) using standard methods of
numerical linear algebra.
%
%
%
\subsection{The GLLS method as a first-order approximation}
\label{sect::gllsm_approx}
%
%
%
If nuclear data uncertainties are sufficiently small, the integral observable
vector $\mb y$ may be approximated by a first-order series expansion about the best-estimate nuclear data vector 
$\bs\alpha_0$:
\begin{align}
\mb y(\bs\alpha) &\,\simeq\, \mb y(\bs\alpha_0) + \mb S\bs\Delta\bs\alpha, \quad
\bs\Delta\bs\alpha \,=\, \bs\alpha-\bs\alpha_0 , \nn\\ 
\left(\mb S\right)_{ij} &\,=\, \left.\partial y_i /\partial\alpha_j\right|_{\bs\alpha=\bs\alpha_0} .
\label{eq::y_first_order}
\end{align}
Choosing a multivariate normal distribution model for the nuclear data vector $\bs\alpha$,
\begin{align}
{\rm p}(\bs\alpha)={\rm N}(\bs\alpha_0,\bs\Sigma_\alpha),
\label{eq::alpha_normal}
\end{align}
the first order expansion yields the following expressions for the distribution model parameters of the prior pdf of $\mb y$:
\begin{align}
\mb y_{0A} &\,\simeq\, \mb y_A(\bs\alpha_0),\quad \mb y_{0B} \,\simeq\, \mb y_B(\bs\alpha_0),\quad
\bs\Sigma_{0A} \,\simeq\, \mb S_A \bs\Sigma_\alpha \mb S_A^T , \nn\\
\bs\Sigma_{0B}  &\,\simeq\, \mb S_B \bs\Sigma_\alpha \mb S_B^T , \quad
\bs\Sigma_{0AB} \,\simeq\, \mb S_A \bs\Sigma_\alpha \mb S_B^T ,
\label{eq::prior_model_first_order}
\end{align}
cf.~Eq.~(\ref{eq::partition_prior_like_post}). Here $\mb S_A$ and $\mb S_B$ denote the application case and benchmark submatrices of 
the sensitivity matrix $\mb S$. 

Inserting Eq.~(\ref{eq::prior_model_first_order}) into Eq.~(\ref{eq::post_model_no_constraints})
yields the well-known GLLS results of the posterior model parameters \citep{cecchini,humi,hemment,broadhead,saintjean,salvatores}:
\begin{align}
\mb y_A^* &\,\simeq\, \mb y_{0A} + \mb S_A \bs\Delta\bs\alpha^* , \quad \mb y_B^* \,\simeq\, \mb y_{0B} + \mb S_B \bs\Delta\bs\alpha^* , 
\nn\\
\bs\Sigma_A^* &\,\simeq\, \mb S_A \bs\Sigma_\alpha^* \mb S_A^T , \quad \bs\Sigma_B^* \,\simeq\, \mb S_B \bs\Sigma_\alpha^* \mb S_B^T , \nn\\
\bs\Sigma_{AB}^* &\,\simeq\, \mb S_A \bs\Sigma_\alpha^* \mb S_B^T .
\label{eq::posterior_model_first_order}
\end{align}
with
\begin{align}
\bs\Delta\bs\alpha^* &\,=\, \bs\alpha^* - \bs\alpha_0 \nn\\ 
& \,=\, \bs\Sigma_\alpha \mb S_B^T \left(\mb S_B\bs\Sigma_\alpha\mb S_B^T + \bs\Sigma_{VB}\right)^{-1}
\left(\mb v_B-\mb y_{0B}\right) , \nn\\
\bs\Sigma_\alpha^* &\,=\, \bs\Sigma_\alpha - \bs\Sigma_\alpha \mb S_B^T \left(\mb S_B\bs\Sigma_\alpha\mb S_B^T + \bs\Sigma_{VB}\right)^{-1}
\mb S_B \bs\Sigma_\alpha .
\label{eq::posterior_nudata_first_order}
\end{align}
Hence, the GLLS estimates are linear transformations of the 
maximum-a-posteriori estimate $\bs\alpha^*$ of the nuclear data vector $\bs\alpha$ and of the corresponding posterior nuclear 
data covariance matrix $\bs\Sigma_\alpha^*$. 

At this point, it is important to stress that this linear relationship between integral observables and nuclear data 
is a result of 
first-order perturbation theory expressed by Eq.~(\ref{eq::y_first_order}). 
Since Eqs.~(\ref{eq::posterior_model_first_order}) and (\ref{eq::posterior_nudata_first_order})
represent a first-order approximation of Eq.~(\ref{eq::post_model_no_constraints}), the GLLS approximation yields acceptable predictions 
only for sufficiently small nuclear data uncertainties \citep{gllsm,rochman_tmc}.
%
%
%
\section{Applications} 
\label{sect::examples}
%
%
%
\subsection{Prediction of the reactivity of a water-moderated fuel assembly}
\label{subsect::toyreactor}
%
%
%
As a first test case of the MOCABA procedure, we choose an exercise from the preliminary Phase IV benchmark 
specification \citep{uacsa_bench} for the 
Expert Group on Uncertainty Analysis for Criticality Safety Analysis (UACSA) of the OECD/NEA Working Party on Nuclear Criticality
Safety (WPNCS). 
This UACSA benchmark was defined with the objective to calculate the Pearson correlations due to system parameter 
uncertainties between the neutron multiplication factors $k_{\rm eff}$ of experiments belonging to the same series 
of criticality safety benchmark experiments and to quantify the impact of these correlations on the prediction of the 
$k_{\rm eff}$ value of an application case; see also \cite{stuke_uacsa} and \cite{behler_uacsa}.

\begin{figure}[ht!]
  \begin{center}
    \includegraphics[width=0.4\textwidth]{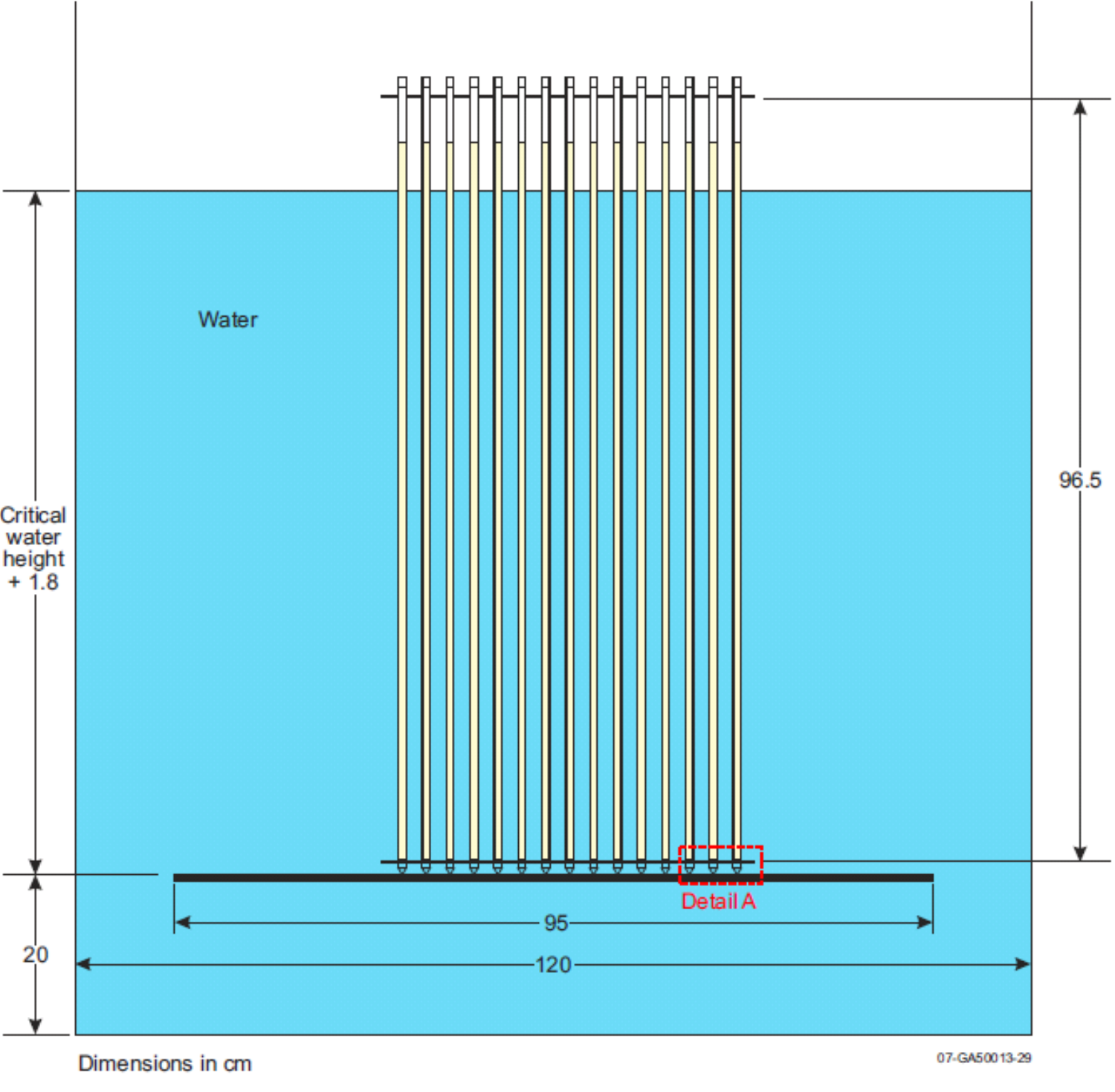}
    \caption[]{\label{fig::lct_config} LCT-007/039: Schematic overview of the experimental setup 
               (adopted from \cite{icsbep}).}
  \end{center}
\end{figure}

The selected exercise involves 21 criticality safety benchmark experiments from the ICSBEP handbook \citep{icsbep}. 
Four experiments are taken from LEU-COMP-THERM-007 (Cases 1, 2, 3 and 4) and 17 experiments from LEU-COMP-THERM-039 
(all 17 cases). Each of these 21 experimental configurations is defined by a single water-moderated array of fuel rods 
in a square pitch arrangement. The fuel rods contained low-enriched $\rm UO_2$ fuel. 
For each experiment, the fuel rod array was placed in a tank and the water level was raised close to the 
critical level; see Figure~\ref{fig::lct_config}. 
The critical water height was then obtained by extrapolation from the subcritical water height 
measurements to the critical water height. 

\red{The 21 configurations differ in the number and pattern of fuel rods within the fuel rod array and in the fuel rod pitch 
which is related to the moderator-to-fuel ratio.}
All experiments made use of the same experimental equipment and the same fuel rods. To cover the stochastic dependencies between 
the $k_{\rm eff}$ values of different criticality experiments, a complete stochastic dependence between the fuel rod 
parameters of different fuel rods is assumed, i.e.~all fuel rods are assumed to be characterized by a single 
parameter for the fuel rod outer diameter, a single parameter for the fuel density, a single parameter for the U-235 enrichment, etc.

Among the benchmark system parameter uncertainties specified in \cite{icsbep}, uncertainties in the fuel rod inner diameter, 
the fuel rod thickness, and the mean linear density of the fuel column 
have the highest impact on $k_{\rm eff}$. Hence, these parameters are chosen as the components of the 
three-dimensional benchmark system parameter vector $\mb x_B$, and the corresponding $3 \times 3$ covariance matrix 
$\bs\Sigma_{XB}$ is derived from the uncertainty specification in \cite{icsbep}. 
The covariance matrix $\bs\Sigma_{VB}$ of the likelihood function is calculated from $\bs\Sigma_{XB}$ 
according to Eq.~(\ref{eq::model_like_linear}),
where the components of the sensitivity matrix ${\mb S_B}$ are derived from SCALE~6.0 criticality calculations for
variations of the three system parameters.

\begin{figure}[hb!]
  \begin{center}
    \includegraphics[angle=0]{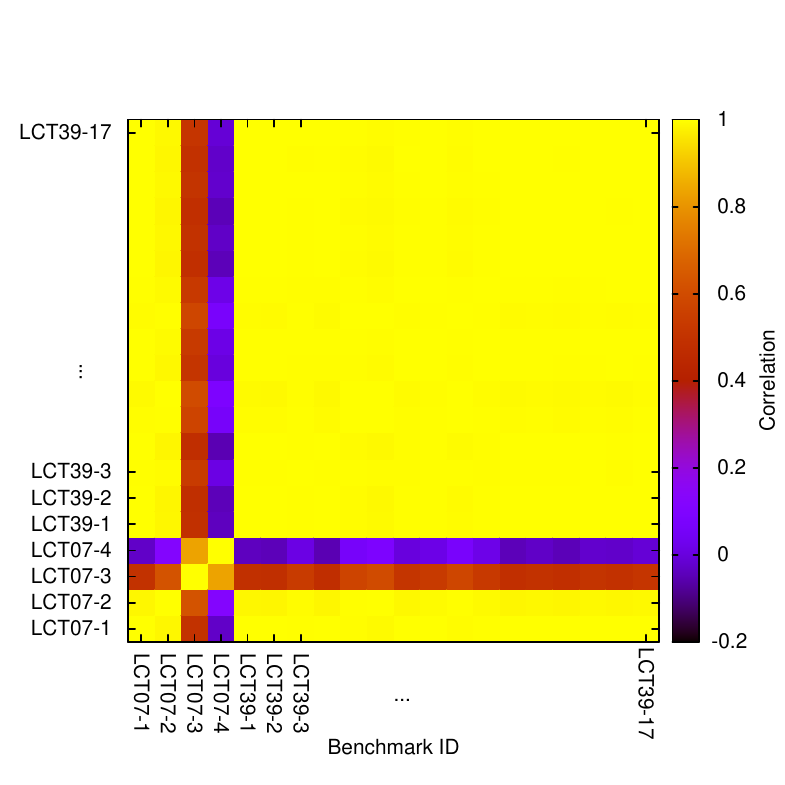}
    \caption[]{\label{fig::Corr_Sigma_B}\red{Pearson correlation matrix corresponding to $\bs\Sigma_{VB}$ for the 21 experiments 
     (Cases 1-4 of LCT-007 and Cases 1-17 of LCT-039).}}
  \end{center}
\end{figure}

As appears from Figure \ref{fig::Corr_Sigma_B}, the Pearson correlations due to system parameter uncertainties between 
the different benchmark $k_{\rm eff}$ values are generally very high. In fact, they are higher than $0.98$ 
except for Case~3 and Case~4 of LCT-007. This is explained by the fact that Case~3 is 
close to optimum moderation and Case~4 is over-moderated, while the remaining 19 experiments are under-moderated. 

\begin{figure}[ht!]
  \begin{center}
    \includegraphics[width=0.3\textwidth]{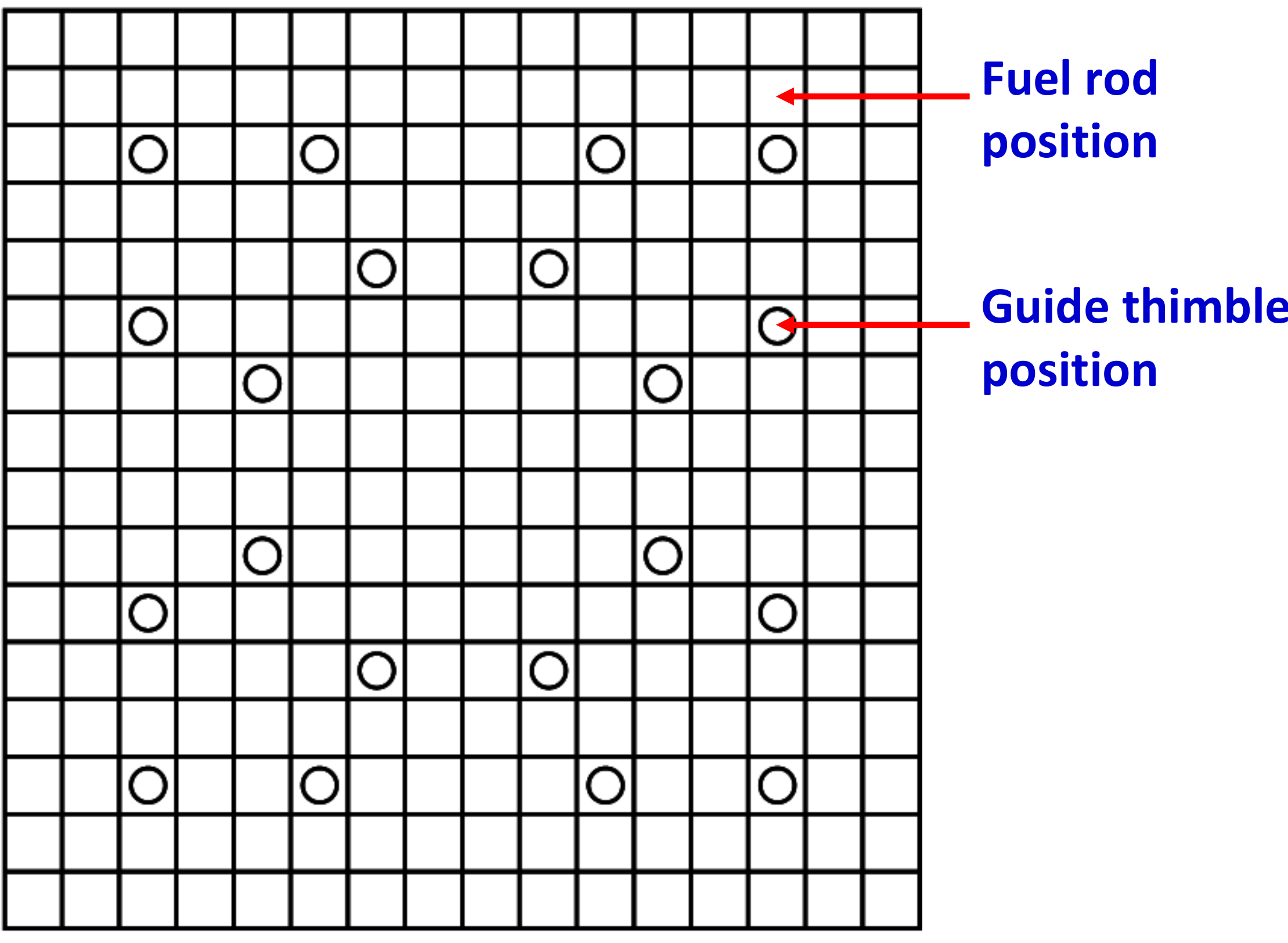}
    \caption[]{\label{fig::16x16_FA}Fuel rod lattice of a $16\times 16$ PWR fuel assembly.}
  \end{center}
\end{figure}

As an application case, we select from \cite{uacsa_bench} the prediction of the 
neutron multiplication factor of a PWR fuel assembly which is moderated and 
fully reflected by pure water; see  Figure~\ref{fig::16x16_FA}. 
Uncertainties are specified for six different system parameters, defining the 
dimensions of the fuel rods, the guide thimbles and the height of the fuel column.
These six parameters are identified with the components of the system parameter vector 
$\mb x_A$; see Section~\ref{sect::model}.
The pdf ${\rm p}(\mb x_A)$ follows directly from the uncertainty specification in \cite{uacsa_bench}.

Following the notation of Section~\ref{sect::model}, the components of the 22-dimensional vector $\mb y$ are given by the 
$k_{\rm eff}$ values of the application case and the 21 benchmark experiments:
\begin{align}
 \mb y &\,=\, \left(y_1,\dots,y_{22}\right)^T \nn\\ 
&\,=\, \left(k_{\rm app},k_{\rm LCT-7-1},\dots,k_{\rm LCT-39-17}\right)^T .
\label{eq::prior_y_uacsa}
\end{align}
For the calculation of the prior and posterior distribution of $\mb y$, we follow the MOCABA procedure 
described in Section \ref{sect::mocaba}. \red{This evaluation is based on 1000 Monte Carlo samples of the nuclear data for $^{235}\rm U$, $^{238}\rm U$, 
$^1 \rm H$ and $^{16}\rm O$ and of the application case system parameters. For the nuclear data sampling the covariance information included 
in the ENDF/B-VII.1 nuclear data evaluation \citep{endfb71} is used.} 
The criticality calculations are performed with SCALE~6.0 using 44-group nuclear data libraries \citep{scale60}.

\begin{figure}[]
  \begin{center}
    \includegraphics[]{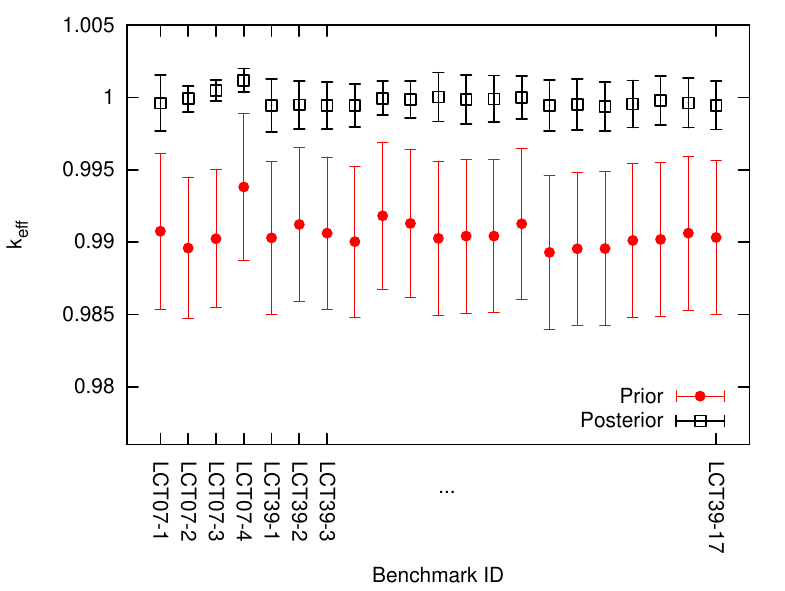}
    \caption[]{\label{fig::prior_post_bench}\red{Mean values of the prior and posterior $k_{\rm eff}$ distributions for the 21 benchmark experiments. The error bars show the standard deviations of the distributions.}}
  \end{center}
\end{figure}
\red{
Figure~\ref{fig::prior_post_bench} shows the mean values for the prior and posterior distributions of the benchmark experiments. One observes a sizable bias in the prior $k_{\rm eff}$ distributions which is predominantly caused by the procedure chosen to collapse the point-wise ENDF/B 7.1 nuclear data to the 44-group SCALE input library. The 44-group libraries provided by ORNL as part of its SCALE~\cite{scale60} package are generated in a two-step process, where first a 238-group library is generated which is then collapsed to 44 groups based on a PWR-like spectrum. In this work, the collapsing was performed in a one-step procedure using a spectrum that represents a combination of a $1/E$, a fission, and a thermal Maxwellian spectrum (NJOY option IWT=4). MOCABA corrects the bias induced by the simplified group collapsing scheme, and thus shifts the posterior distribution towards larger $k_{\rm eff}$ values. More recent NUDUNA applications are based on 238-group libraries and show for low-enriched UO$_{2}$ and MOX lattices moderated by water similar prior biases as default SCALE libraries.}

The evaluation of the prior covariance matrix $\bs\Sigma_0$ according to Eq.~(\ref{eq::MC_estimators}) reveals high correlations due to 
nuclear data uncertainties between the $k_{\rm eff}$ values of the benchmark experiments and of the application case. 
In fact, these correlations are higher than $0.97$ except for Case~3 and Case~4 of LCT-007. 
Hence, we may expect that our knowledge about the application case $k_{\rm eff}$ value will be significantly improved by taking 
into account the benchmark experiments. 
\begin{figure}[ht!]
  \begin{center}
    \includegraphics[width=0.4\textwidth]{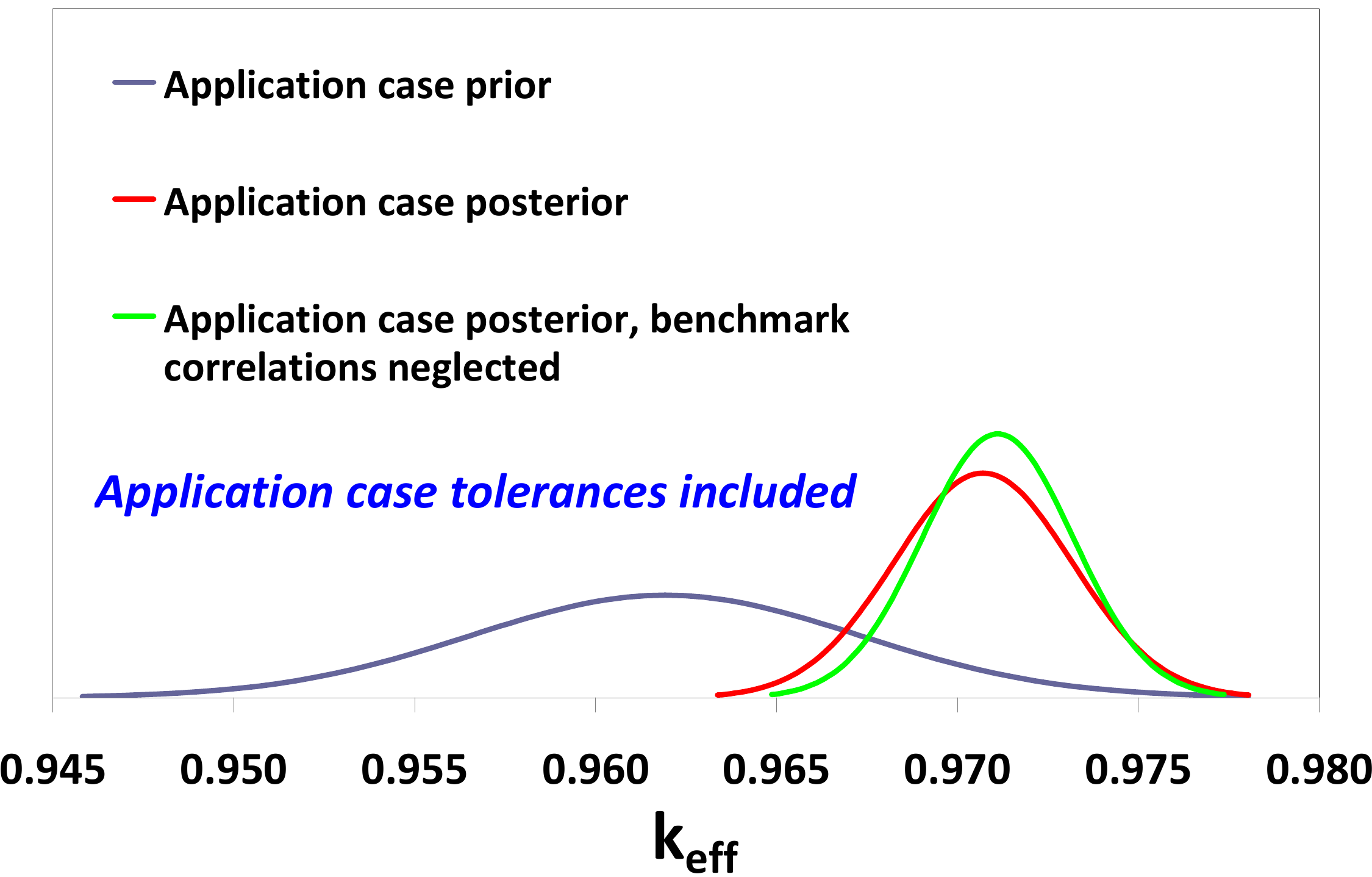}
    \includegraphics[width=0.396\textwidth]{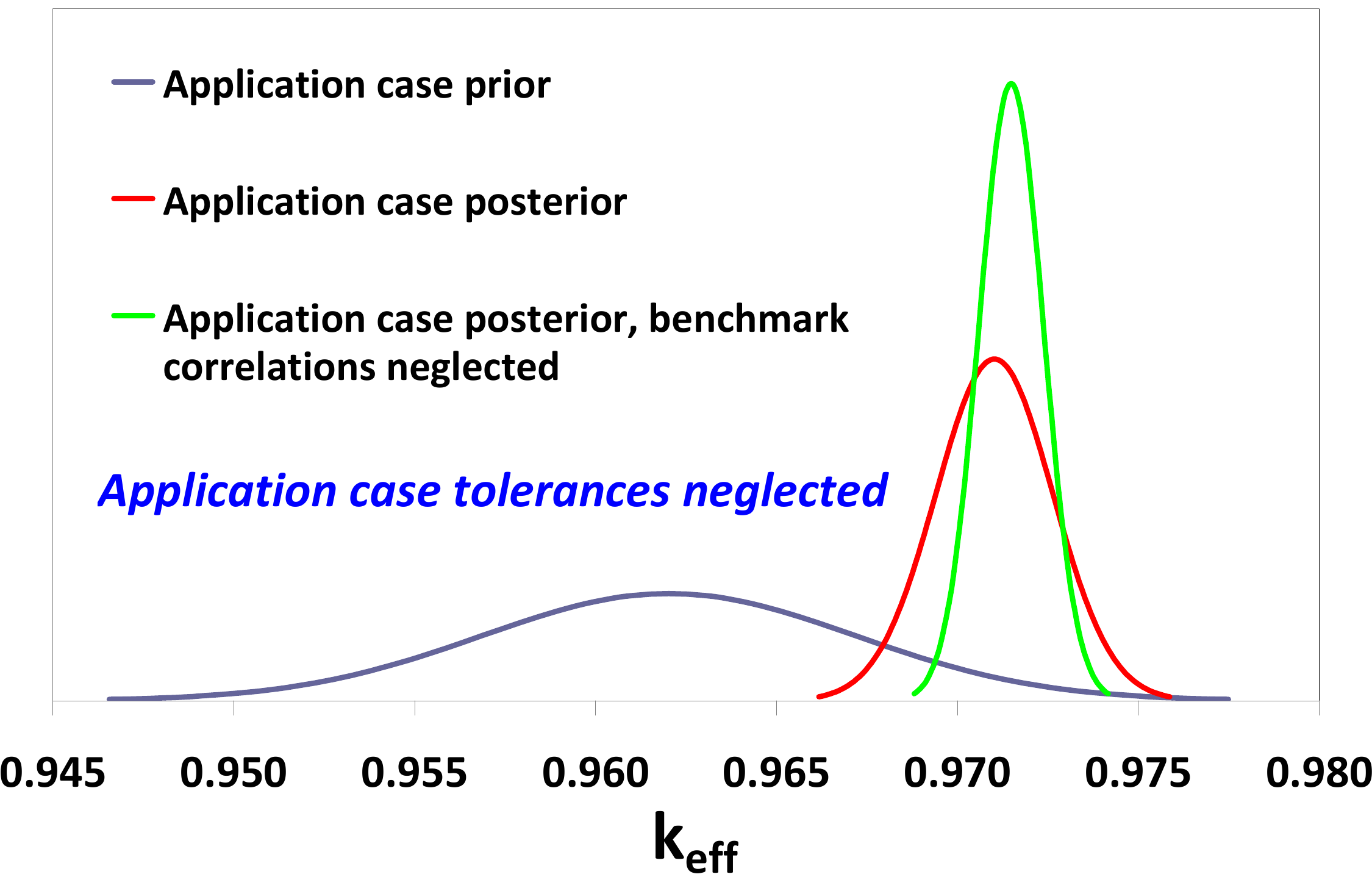}
    \caption[]{\label{fig::prior_post_app}Prior and posterior $k_{\rm eff}$ distribution for the application case. 
    For the lower panel, the system parameter uncertainties of the application case were neglected, i.e.~the application case system 
    parameter vector $\mb x_A$ was kept constant at its nominal value (no Monte Carlo sampling of $\mb x_A$, see Section~\ref{sect::mocaba}).}
  \end{center}
\end{figure}
\begin{figure}[ht!]
  \begin{center}
    \includegraphics[]{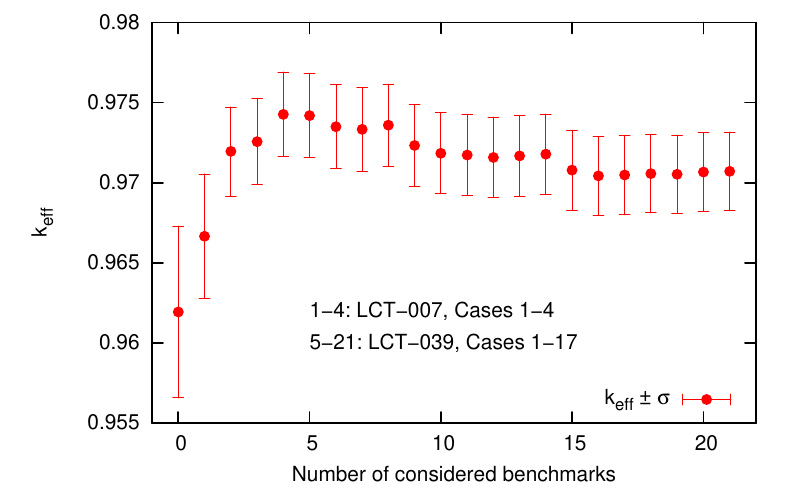}
    \caption[]{\label{fig::post_comvergence}\red{Posterior $k_{\rm eff}\pm \sigma$ for the application case as a function of the number of benchmark
    experiments taken into account in the Bayesian updating procedure.}}
  \end{center}
\end{figure}

This expectation is confirmed by Figure~\ref{fig::prior_post_app}, since the width of the posterior distribution is significantly reduced 
compared to the width of the prior distribution. As also appears from this figure, 
covariances due to system parameter uncertainties between the $k_{\rm eff}$ values of different benchmark experiments 
enhance the width of the posterior $k_{\rm eff}$ distribution which reflects a decrease in information. 
Hence, it is generally important to take covariances due to system parameter uncertainties into account. 

\red{Figure~\ref{fig::post_comvergence} shows the posterior $k_{\rm eff}\pm \sigma$ values of the application case as a function of the number of benchmark 
experiments taken into account. As can be seen, the posterior $k_{\rm eff}$ predictions become nearly stable after the first five benchmark 
experiments have been added, and the posterior uncertainty is not further reduced. The reason is, as follows from Figure~\ref{fig::prior_post_app}, 
that after adding the first 5 benchmark experiments the posterior uncertainty is dominated by the application case tolerances, and this uncertainty 
cannot be reduced by adding more benchmark information.}
%
%
%
\subsection{Power distribution of a toy model reactor}
\label{subsect::toyreactor}
%
%
%
The example presented in the last sub-section was related to the prediction of a single integral observable, namely the
neutron multiplication factor of a water-moderated fuel assembly, where the integral observable vector $\mb y_A$
of the application case had only one component. Since MOCABA may be applied to the prediction 
of any vector function of a nuclear data vector $\bs\alpha$, we are now going to test the updating algorithm of MOCABA 
for a simple reactor toy model where $\mb y_A$ has dimension 100. The purpose of this toy model, which is similar to the one used for an 
exercise in \cite{uacsa_bench}, is not to simulate a real reactor core but to demonstrate the MOCABA method 
\red{for a high-dimensional integral vector function}.

Each component of $\mb y_A$ may be thought of as the thermal power of one out of 100 fuel assemblies within a nuclear 
reactor. Hence, the sum of the components of $\mb y_A$ represents the total reactor power. 

We assume a four-dimensional nuclear data random vector $\bs\alpha$ defined by the following pdf:
\begin{align}
{\rm p}(\bs\alpha) &\,=\, {\rm N}(\bs\alpha_0,\bs\Sigma_\alpha), \nn\\
\bs\alpha_0 &\,=\,\left(9.9968,1.0066,1.0225,1.2198 \right)^T, \nn\\
\bs\Sigma_\alpha &\,=\, \mb{diag} \left(0.3,0.03,0.03,0.03\right) \,.
\label{eq::toy_p_alfa}
\end{align}
$\mb y_A$ is supposed to be given by the following vector function:
\begin{align}
 \mb y_A &\,=\, (y_{A,1},\dots,y_{A,100})^T ,  \quad
  y_{A,i} \,=\, \bs\alpha^T \mb x_{A,i}, \nn\\
 {\rm p}(\mb x_{A,i}) &\,=\, {\rm N}\left(\mb x_{A,0},\mb\Sigma_{XA}\right), \quad
 \mb x_{A,0} \,=\, \left(0.6,6,-12,24 \right)^T, \nn\\ 
\mb\Sigma_{XA} &\,=\, \mb{diag} \left(0.006,0.06,0.12,0.24\right).
 \label{eq::toy_y_A}
\end{align}
We further assume nine benchmark measurements being related to the following vector function:  
\begin{align}
 \mb y_B &\,=\, (y_{B,1},\dots,y_{B,9})^T , \nn\\
 y_{B,i} &\,=\, \frac{\alpha_1\alpha_4 x_{B1}}{\alpha_1x_{B1}+\alpha_2x_{B2}^{(i)}+\alpha_3x_{B3}^{(i)}} .
 \label{eq::toy_y_B}
\end{align}
The components $y_{B,i}$ of $\mb y_B$ may be thought of as neutron multiplication factors of nine different critical configurations,
i.e.~the measurements of $y_{B,i}$ are given by $v_{B,i}=1$, and the uncertainty distributions of the benchmark system parameters 
follow directly from Table~\ref{tab::sys_9bench}, where we assume normal distribution models for all system parameters.

\begin{table*}[ht]
  \begin{center}
    \caption{\label{tab::sys_9bench}System parameters defining the 9 benchmark experiments.}
    \footnotesize
    \begin{tabular}{cllllllll}
    \toprule
    Benchmark i & $x_{B1}$ & $\sigma_{B1}$ & $x_{B2}^{(i)}$ & $\sigma_{B2}^{(i)}$ & $x_{B3}^{(i)}$ & $\sigma_{B3}^{(i)}$ & $y_{B,i}^{(0)}$ & 
              $v_{B,i}$ \\
    \midrule
    1 & 2.0072   & 0.05         &  4.0424        & 0.05              & -0.0746       & 0.05              & 1.0174   & 1.0 \\
    2 & 2.0072   & 0.05         &  1.9601        & 0.05              &  1.9292       & 0.05              & 1.0194   & 1.0 \\  
    3 & 2.0072   & 0.05         & -0.0506        & 0.05              &  3.9477       & 0.05              & 1.0177   & 1.0 \\ 
    4 & 2.0072   & 0.05         & -2.0458        & 0.05              &  6.0650       & 0.05              & 1.0111   & 1.0 \\
    5 & 2.0072   & 0.05         & -3.9905        & 0.05              &  8.0370       & 0.05              & 1.0086   & 1.0 \\  
    6 & 2.0072   & 0.05         & -6.0613        & 0.05              &  9.8448       & 0.05              & 1.0185   & 1.0 \\ 
    7 & 2.0072   & 0.05         & -12.0059       & 0.05              & 15.9819       & 0.05              & 1.0063   & 1.0 \\
    8 & 2.0072   & 0.05         & -16.0923       & 0.05              & 19.9995       & 0.05              & 1.0066   & 1.0 \\  
    9 & 2.0072   & 0.05         & -20.0440       & 0.05              & 23.9692       & 0.05              & 1.0032   & 1.0 \\ 
    \bottomrule
    \end{tabular}
  \end{center}
\end{table*}

The second to last column in Table~\ref{tab::sys_9bench} contains the calculated benchmark $k_{\rm eff}$ values $y_{B,i}^{(0)}$ 
for the nominal values of the nuclear data and the benchmark system parameters.

The observed Pearson correlations due to nuclear data uncertainties between the elements of the application case vector $\mb y_A$ and 
the benchmark vector $\mb y_B$, corresponding to the submatrix $\bs\Sigma_{0AB}$ of $\bs\Sigma_0$ 
(see Eq.~(\ref{eq::partition_prior_like_post})), are in the range between $0.7$ and $0.93$. 
Hence, we may expect that taking into account the benchmark measurements will significantly improve our knowledge about $\mb y_A$.
\begin{figure}[ht!]
  \begin{center}
    \includegraphics[width=0.4\textwidth]{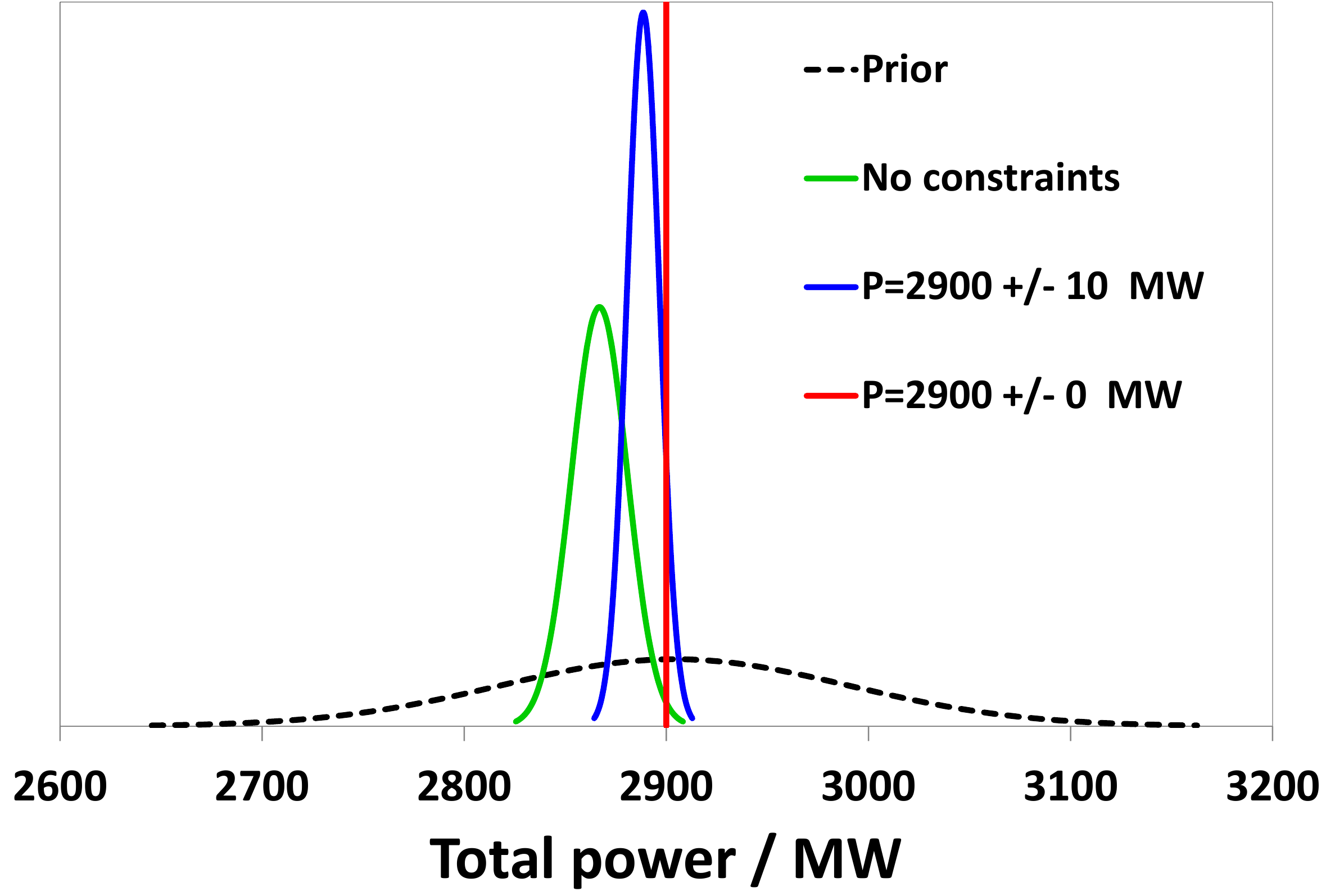}
    \includegraphics[width=0.4\textwidth]{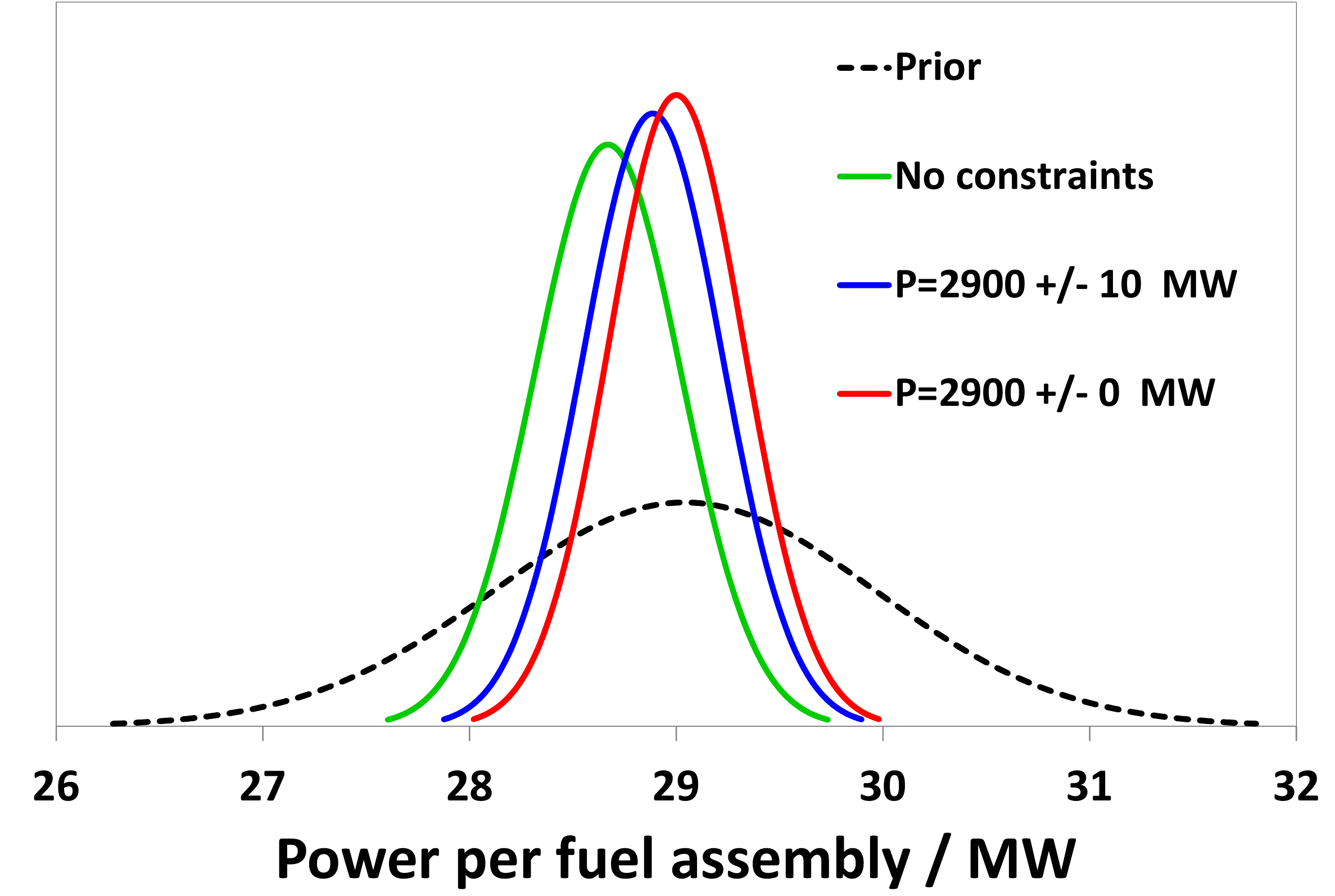}
    \caption[]{\label{fig::toy_power}Prior and posterior distributions for total reactor power and power per fuel assembly.}
  \end{center}
\end{figure}

This expectation is confirmed by Figure~\ref{fig::toy_power}, which shows the prior and posterior pdfs of the total reactor
power, i.e.~the sum of all 100 components of $\mb y_A$, and of the power $y_{A,i}$ of a single fuel assembly. As can be seen, 
the widths of the posterior distributions are much smaller than those of the prior distributions, which reflects 
a strong increase in information. Here the posterior distributions are presented for three different 
cases: the green curves correspond to the case where no constraints are imposed on the total reactor power, 
the blue curves correspond to a total power of $2900~\rm MW$ with a standard deviation of $10~\rm MW$, and the red curves
correspond to a total power of exactly $2900~\rm MW$.\footnote{\red{It should be noted that for a real reactor power simulation the 
total power is fixed, i.e.~the power distribution is normalized to the total power. This means that the sum constraint has to be applied 
already to the prior local power vector $\mb y$ before taking into account any measurements. 
According to Section~\ref{sect::model}, within the MOCABA framework this sum constraint
(as a special case of a linear constraint) is imposed by means of Bayesian updating of the unconstrained prior pdf of $\mb y$. 
Information from benchmark measurements can then be included in a second updating step. For the considered toy model this two-step updating 
procedure, which is mathematically equivalent to a single combined updating step, yields the red curves shown in 
Figure~\ref{fig::toy_power} if the total power is constrained to be 2900 MW.}}

Predicting the power distribution for a real reactor core would follow \red{a similar} procedure
as for the toy model example, except that one would use NUDUNA for the nuclear data sampling, and the analytic expressions in 
Eqs.~(\ref{eq::toy_y_A}) and (\ref{eq::toy_y_B}) would be replaced by transport calculation results.
%
%
%
\section{Conclusions} 
\label{sect::conclusion}
MOCABA combines the advantages of Monte Carlo based nuclear data uncertainty propagation 
(no first order approximation, no adjoint calculations necessary, applicable to any 
function of nuclear data) with those of the Generalized Linear Least Squares method (updating of predictions of
integral observables by taking into account integral measurements). 
Being based on a general Bayesian scheme, MOCABA can be applied to the prediction of any kind of integral 
observable (neutron multiplication factors, isotopic concentrations in irradiated nuclear fuel, 
reactor power distributions, etc.), and any integral measurement may be used as a benchmark to update
the prediction of an integral observable. 
Additionally, constraints on linear combinations of integral observables can be accounted for, e.g.~a constraint on 
the total reactor power for the prediction of a reactor power distribution. MOCABA incorporates the NUDUNA code for the 
nuclear data sampling. Since NUDUNA performs its sampling directly on evaluated nuclear data files in ENDF-6 format, 
the MOCABA procedure is not restricted to any particular nuclear data library format of a transport code system. 
The current NUDUNA version supports automatic compilation of ACE and AMPX libraries, i.e.~MOCABA can already be used 
in combination with continuous energy MCNP or SERPENT transport calculations and multigroup SCALE calculations. 
A NUDUNA upgrade for generating few-group libraries for reactor core simulation systems is in development, so that the 
MOCABA code system will also be applicable to reactor core design and reactor safety analysis.

%
%
%
%
%





\section*{Acknowledgements}
This work was conducted in the framework of the  \linebreak AREVA~GmbH R\&D project "Uncertainty Analysis and Uncertainty Propagation in Nuclear Design Systems"'. 
We thank M.~Lamm for his support in R\&D project management.


\bibliographystyle{elsarticle-harv}
\bibliography{mocaba_ann_nuc}







\end{document}